\documentstyle[11pt,newpasp,twoside,epsf]{article} 
\def\edcomment#1{\iffalse\marginpar{\raggedright\sl#1\/}\else\relax\fi}
\reversemarginpar

\begin{document}
\title{A Far-Ultraviolet Spectroscopic Survey of the Globular Cluster
47~Tuc} 
\author{Christian Knigge} 
\affil{University of Southampton, Department of Physics and Astronomy,
Highfield, Southampton, SO17 1BJ, UK} 
\author{Michael M. Shara \& David R. Zurek} 
\affil{American Museum of Natural History, Department of Astrophysics,
Central Park West, New York, NY 10024, USA} 
\author{Knox S. Long \& Ronald L. Gilliland}
\affil{Space Telescope Science Institute, 3700 San Martin Drive,
Baltimore, MD 21218, USA}

\begin{abstract}
We present preliminary results from a FUV spectroscopic and photometric
survey of the globular cluster 47~Tuc. Our main goal is to either
confirm or rule out the existence of a large population of CVs in the
core of this cluster. 

We have so far identified approximately 425 FUV sources, most of which
are probably hot, young WDs. However, we have also found approximately
30 FUV sources whose position in a FUV-optical color-magnitude diagram
makes them strong CV candidates. If most or all of these objects are
eventually confirmed as CVs, the large CV population predicted by
tidal capture theory has finally been found.

Our data has also allowed us to resolve the long-standing
puzzle surrounding AKO~9, a UV-bright and highly variable 1.1~day
binary system in 47~Tuc. AKO~9 is the brightest FUV source in our
data and presents a blue FUV spectrum with strong C~{\sc iv} and
He~{\sc ii} emission lines. Its spectrum is similar to that
of the long-period, DN-type field CV GK~Per. Based on this similarity
and other evidence, we suggest that AKO~9 is a long-period
CV in which mass-transfer is driven by the nuclear evolution of a
sub-giant donor star.
\end{abstract}

\section{Introduction}

Globular clusters (GCs) are fantastic stellar crash test
laboratories. Because of the huge number of stars that are crammed
into the relatively small core of a GC, close encounters between
cluster members are almost ubiquitous. It has been estimated that up
to 40\%~of the stars in the cores of some GCs have suffered a physical
collision during their lifetime (Hills \& Day 1976).  Near misses are
even more common and are thought to produce the close binary systems
(CBs) that in turn can drive the dynamical evolution of entire GCs
(Hut et al. 1992).

The dominant mechanism for CB production in GCs is thought to be
``tidal capture'' (Fabian, Pringle \& Rees 1975). This term describes
the outcome of a close encounter between a degenerate object (neutron
star [NS] or white dwarf [WD]) and a main-sequence (MS) star. In the
course of such an encounter, strong tides are raised on the surface of
the MS star. These distortions dissipate orbital energy and can
therefore result in capture of the MS star and CB formation.
 
Tidal capture theory was developed to account for the
overabundance of x-ray sources (thought to be NS/MS binaries) in the
cores of GCs (e.g. Katz 1975). However, tidal capture theory also
predicts that even more cataclysmic variables (CVs) -- interacting
WD/MS binary systems -- should exist in GC cores. In fact, detailed
simulations predict the presence of well over 100 CVs in each of
47~Tuc and $\omega$~Cen (Di Stefano \& Rappaport 1994). 
Despite several HST-based searches for these systems in several
clusters, only a handful of candidates have so far been found 
(Table~1). If this apparent dearth of CVs is real, it raises 
serious questions about tidal capture theory and perhaps even about
our binary-driven view of GC evolution. 

\begin{table}[h] 
\caption{Candidate Cataclysmic Variables in Globular Clusters}
\begin{center}
\begin{tabular}{ccc}
&&\\ \hline 
\multicolumn{2}{c}{Globular Cluster} & Known CV \\ 
NGC Number & Other Name & candidates \\ \hline
NGC 6397 &  		& 3 	\\
NGC 6681 &  		& 0 	\\
NGC 7078 & M 15 	& 0 	\\
NGC 6441 &  		& 1 	\\
NGC 6624 &  		& 1 	\\
NGC 6293 &  		& 0 	\\
NGC 1851 &  		& 0 	\\
NGC 7099 & M 30 	& 0 	\\
NGC 6752 &  		& 2-4 	\\
NGC 104~ & 47 Tuc 	& 3  	\\
NGC 6093 & M 80 	& 3 	\\
NGC 5904 & M 5 		& 1 	\\
NGC 5927 &  		& 0 	\\
NGC 6637 & M 69 	& 0 	\\
NGC 6402 & M 14 	& 1 	\\
NGC 6171 & M 107	& 0 	\\
NGC 6352 &  		& 0 	\\ \hline
\end{tabular}
\end{center}
\end{table}

Unfortunately, none of the previous searches for CV in GCs has been
definitive. The candidates found so far typically have absolute
magnitudes $M_V \sim 7$, which is close to the effective detection
limit for CVs in these surveys. The majority of CVs in the field are
at least 1-2 mags fainter than this (Figure~1). Most of these objects
could simply not have been detected in GCs to date.  Moreover, all
previous searches identified CV candidates on the basis of just one or
two characteristics (e.g. variability due to orbital motion, dwarf
nova [DN] eruptions, H$\alpha$~emission, UV excess). But not all field
CVs exhibit all of these characteristics: low inclination systems
display little or no orbital variability; most magnetic CVs do not undergo
DN outbursts; and some disk-accreting CVs exhibit very weak H$\alpha$
lines. Sensitivity to many distinguishing characteristics is therefore
vital if a CV census is to be reliable. Previous searches have not
been ``broad'' enough in this sense. It is therefore unclear whether
the large number of CVs predicted by tidal capture theory exist in the
cores of GCs.

\begin{figure}
\plotfiddle{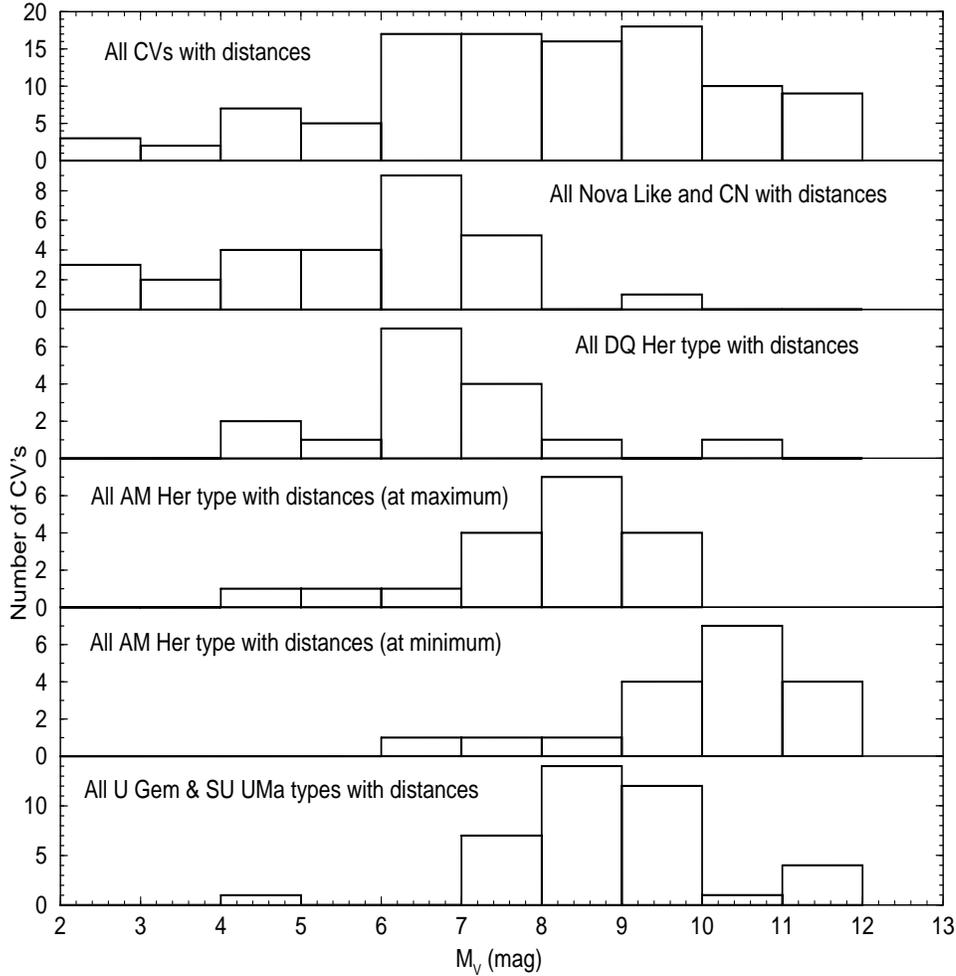}{350pt}{270.01}{55}{65}{-215}{390}
\caption{The absolute magnitude distributions of field CVs with 
known distances, broken down by CV sub-type. From top to bottom, we
show the distributions for all CVs, non-magnetic nova-likes and
classical novae, intermediate polars, polars at maximum, polars at
minimum, and  dwarf novae in quiescence.}
\end{figure}

\section{Observations}

In order to obtain a definitive, empirical answer to this question, we
are currently carrying out a new HST-based search for GC CVs. However,
our search is based on a completely novel observational strategy. Its 
two key components are: (i) a strong emphasis on the far-ultraviolet (FUV)
waveband (1450\AA-1750\AA); (ii) the combination of FUV and optical
photometry with slitless, multi-object FUV spectroscopy. 
We can therefore achieve immediate spectroscopic confirmation of 
most photometric CV candidates. Our program also provides information
on stellar variability a wide range of time-scales, from 10~mins
(individual exposures), to hours (single HST visits), to weeks and
months (spacing between HST visits).

Our survey is targetted on the well-known cluster 47~Tuc, whose
proximity ($m-M = 13.3$), low reddening ($E_{B-V} = 0.04$) and high
metallicity ([Fe/H=-0.76]) makes it ideal for our purposes. (The
high metallicity is important because it means there are no blue
horizontal branch stars that may swamp the FUV light of candidate
CVs.)  We have observed 47~Tuc with the {\em Space Telescope Imaging
Spectrograph} (STIS) for a total of 30 HST orbits, spread
over 6 visits of 5 orbits each. One pilot visit occurred in
September 1999; the remainder took place during a two week window
in August 2000. The majority of time in each visit was spent on
slitless FUV spectroscopy with the G140L grating, implemented as a
series of successive 10~min exposures. A quartz filter was used to
block all light shortward of about 1450~\AA~in order to suppress the
strong geocoronal 
backgrounds due to Ly$\alpha$ and O~{\sc i}.  The remaining time in
each visit was spent on a sequence of 10~min FUV images and 1~min
optical images. The FUV images are important for several reasons: (i)
they allow us to search for short time-scale variability essentially
all the way down to the spectroscopic detection limit; (ii) their sum
actually goes deeper than our FUV spectroscopy; (iii) they allow us to
assess and correct for any remaining crowding in the slitless FUV
spectral images. The main purpose of the optical images is to tie our
data to existing HST observations of 47~Tuc, including the extremely
deep WFPC2 data set that has been obtained as part of a recent search
for planets in this cluster (Gilliland et al. 2000).

Our CV search is both deeper and broader (in the sense defined above)
than any previous surveys. More specifically, the crucial
characteristics of our program are:

\noindent {(i) Depth: We expect to detect faint CVs down to $M_V
\simeq 11$ spectroscopically and $M_V \simeq 13$ photometrically (in
our FUV images). Thus our census of CV candidates for the observed
part of this cluster will be essentially complete
(c.f. Figure~1). Taking all observational limiting factors into
account, we find that Di Stefano \& Rappaport's (1994) simulations
provide a firm target: we should discover and confirm at least 25 CVs
spectroscopically and detect an additional 15 in our FUV images.

\noindent {(ii) Breadth:} Our FUV search is sensitive to {\em all} of
the following CV characteristics: UV brightness, blue FUV spectral
shape, strong C~{\sc iv}~1550~\AA and He~{\sc ii}~1640~\AA~lines,
flickering variability (on time-scales of minutes), orbital
variability (on time-scales of hours), DN eruptions and high/low state
variability (on time-scales of weeks to months).

\section{Results}

The analysis of the observations is still ongoing. The results
presented below are therefore preliminary and based only on the data
obtained during the September 1999 pilot visit. 

\subsection{White Dwarfs, Blue Stragglers and Cataclysmic Variables}

Figure~2 shows a comparison of the FUV and optical images of the
cluster core. Note the dramatic difference in crowding between the two
exposures. The positions and error circles of x-ray sources, blue
stragglers (BSs) and other unusual objects known previously are
marked in the FUV image. There are several clear coincidences, but
most of our FUV sources do not have a well-known optical or x-ray
counterpart.

\begin{figure}
\plotfiddle{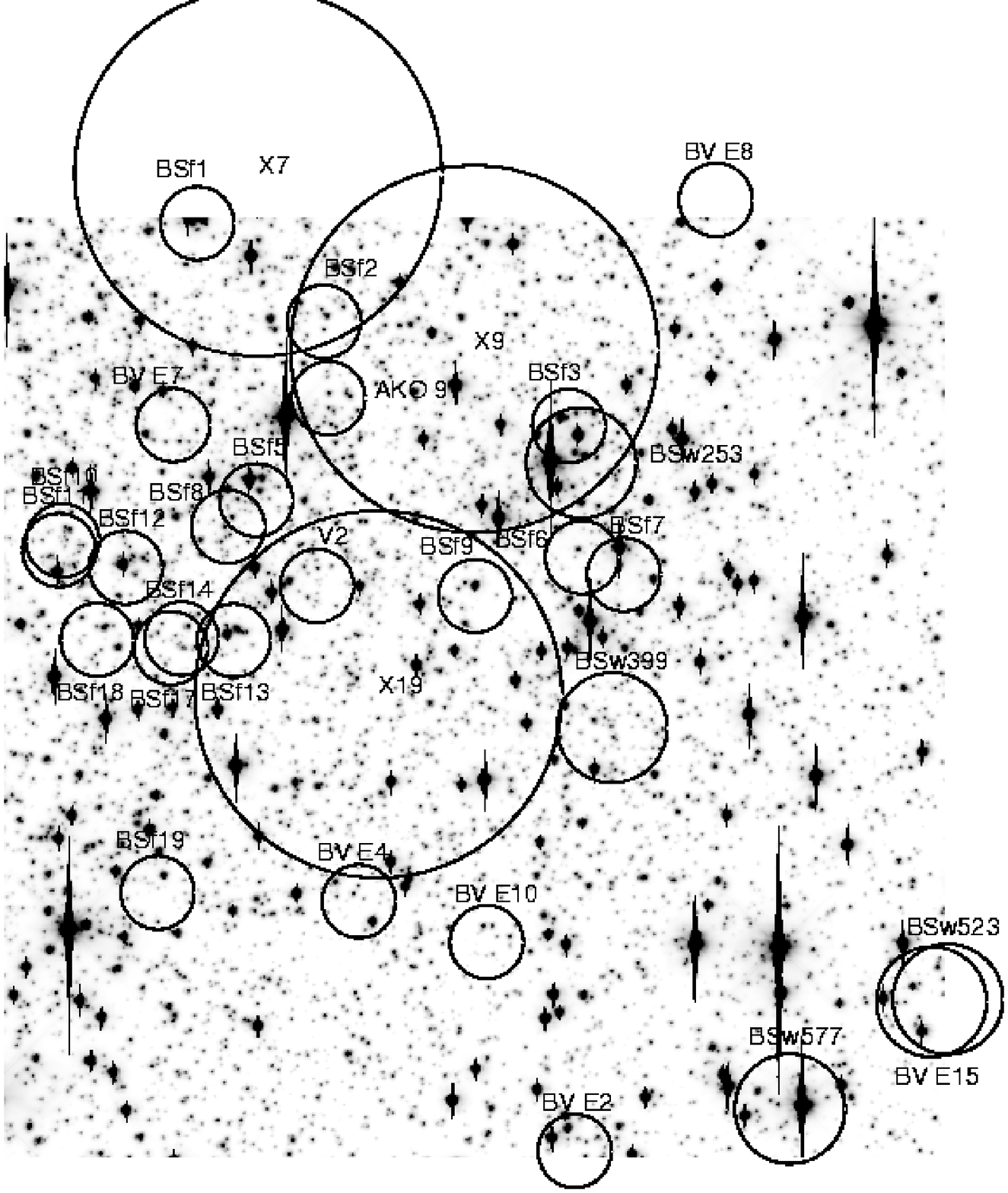}{260pt}{0}{40}{40}{-100}{-35}
\plotfiddle{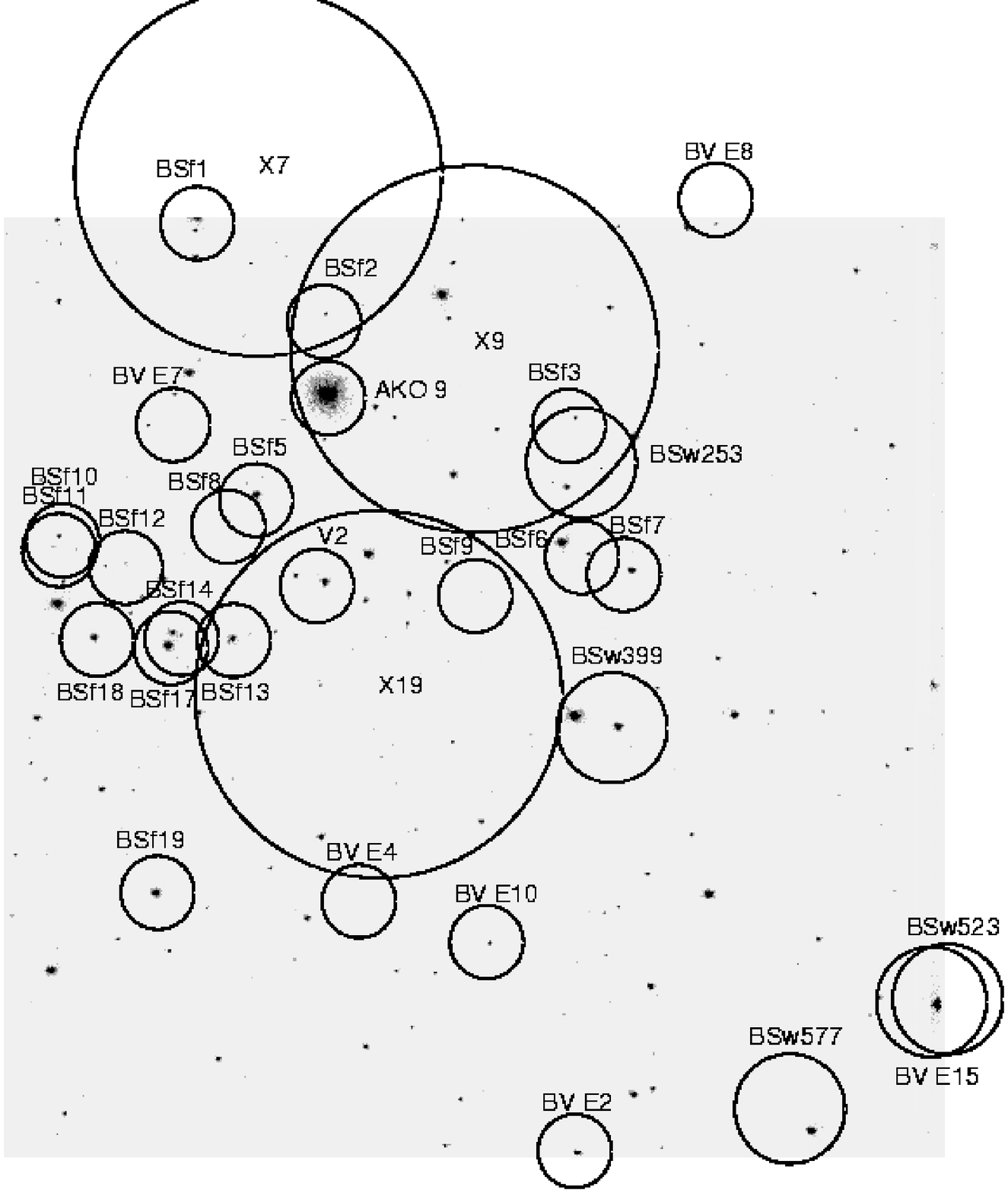}{260pt}{0}{40}{40}{-100}{-35}
\caption{A comparison of optical (top panel) and FUV (bottom panel)
images of the core of 47~Tuc. The intrinsically larger CCD image has
been cropped to cover the same field-of-view as the FUV exposure. 
Note the extreme difference in crowding:
main sequence stars, red giants and red horizontal branch stars are
too cool to show up in the FUV. The positions (with error circles) of 
the following objects are marked: V2 = dwarf nova (Shara et
al. 1996); AKO9 = variable blue object (Auri\protect{\`{e}}re, Koch-Miramond \&
Ortolani (1989; see also 3.2); X = x-ray sources from Hasinger et
al. (1994); BSf=blue stragglers discovered in HST/FOC data (Paresce
et al. 1991); BSw =  blue stragglers discovered in HST/WFPC data
(Guhathakurta et al. 1992; Gilliland et al. 1995); BVE = blue
variables from Edmonds et al. (1996).}
\end{figure}

Figure~3 shows the summed FUV spectral image of the cluster core, as
produced by our slitless spectroscopy. Each trail in this figure is
the dispersed image of a bright FUV point source. The sharp cut-off at
the left hand side of each trail corresponds to the abrupt decrease in
sensitivity shortward of 1450~\AA, where the quartz filter becomes
opaque. This cut-off can actually be used to provide a ``rough-and-ready''
wavelength calibration. The spectra of well isolated, bright objects
can be extracted fairly straightforwardly from the spectral image,
using a method analogous to aperture photometry (i.e. by placing a
'virtual' slit on the image). However, the extraction of overlapping
and/or faint spectral trails requires the spectral equivalent of
PSF-fitting photometry.

\begin{figure}
\plotfiddle{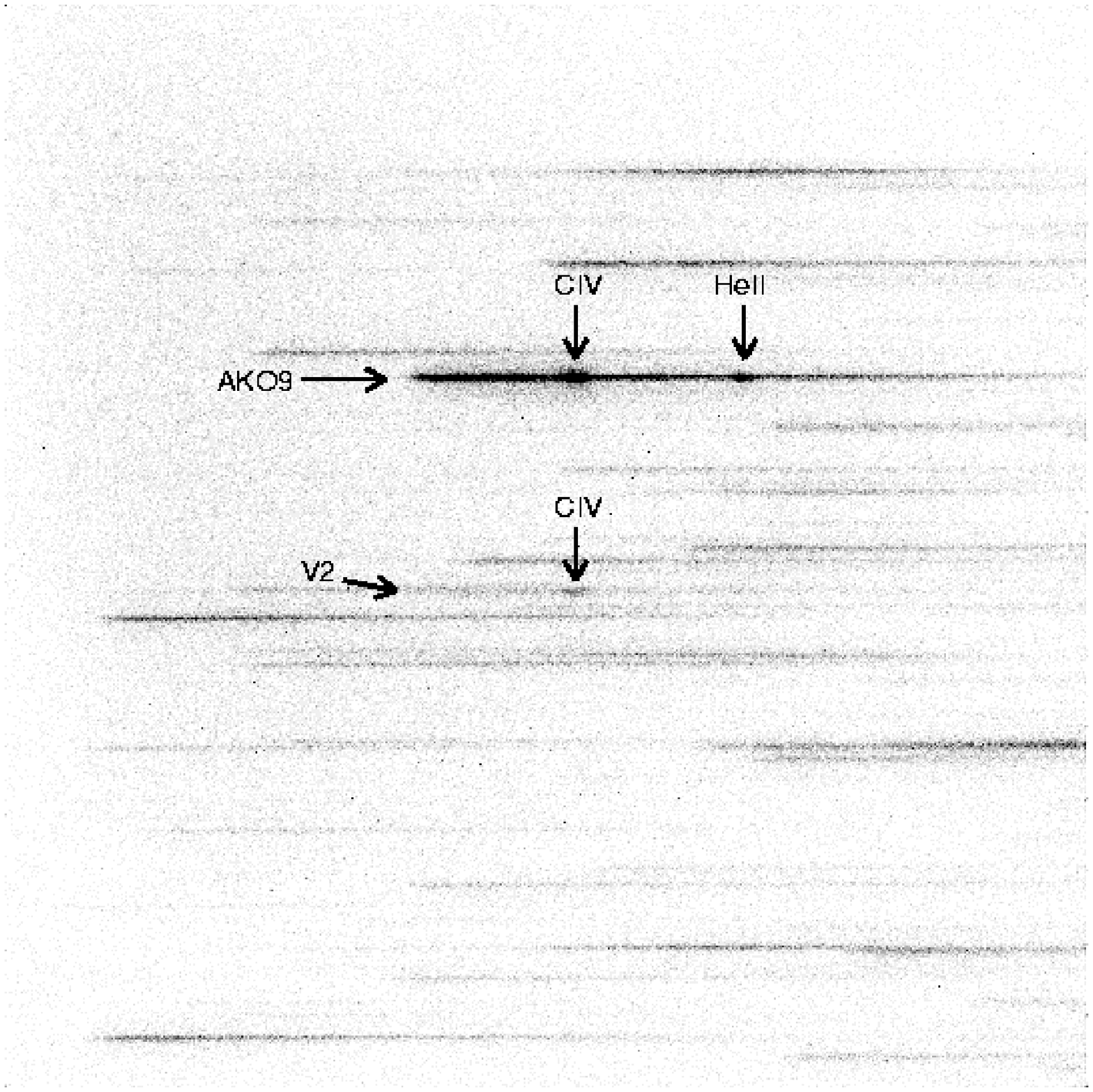}{500pt}{0}{70}{70}{-205}{-35}
\caption{The summed FUV spectral image produced by our slitless
spectroscopy. The two obvious emission line sources -- AKO~9 and V2 --
are marked. We used STIS in its FUV-MAMA/F25QTZ/G140L configuration,
which normally covers a wavelength range of about 1450~\AA~
1720~\AA~at 0.6~\AA~resolution. However, because sources are not
centered on the detector, some spectra actually extend to 
longer wavelengths. Note that there is a vertical offset between
the FUV images and spectra. This offset is automatically applied to all
STIS/MAMA spectroscopy and causes approximately 15\% data loss in our
spectra since the top part of the detector is not actually exposed to
the sky (note the absence of spectral trails in the top of the
spectral image).}
\end{figure}

Figure~4 shows the FUV luminosity function, which rises fairly steeply
towards fainter magnitudes. By far the brightest FUV source in our
images is the well-known object AKO~9, to which we will return in
Section~3.2. The total number of FUV sources we have 
detected to date is $\simeq$425. Most of these are likely to be WDs,
as shown by the following simple estimate (c.f. Richer et al. 1997) 
The number of stars in two post-main-sequence
phases is, in general, proportional to the duration of these
phases. We will use horizontal branch (HB) stars as a reference
point, and therefore take $N_{HB}$ as the number of HB stars in our
field of view (FOV) and
$\tau_{\sc hb} \simeq 10^8$~yrs as their lifetime (e.g. Dorman 1992). We
then expect to find approximately
\begin{equation}
N_{WD} (T_{eff} > T_{lim}) \simeq N_{HB} \times
\left[\frac{\tau_{\sc wd}(T_{lim})}{\tau_{\sc hb}}\right],
\end{equation}
WDs hotter than $T_{lim}$ in the FOV, where 
$\tau_{\sc wd}(T_{lim})$ is the time it takes a WD to cool to
$T_{lim}$. Inspection of existing 
optical data shows that $N_{HB} \simeq 50$ (Guhathakurta et
al. 1992). The appropriate WD cooling time can be estimated from the
fact that we can detect WDs down to approximately $T_{lim} \simeq
9000$~K in our summed FUV image. WD models predict a cooling time of 
$\tau_{\sc wd} \simeq 7-8 \times 10^8$~yrs to reach this temperature
(e.g. Wood 1995). Substituting these numbers into Equation~1, we find
that we may expect to detect $N_{WD} \simeq 350-400$ WDs in our FOV.

\begin{figure}
\plotfiddle{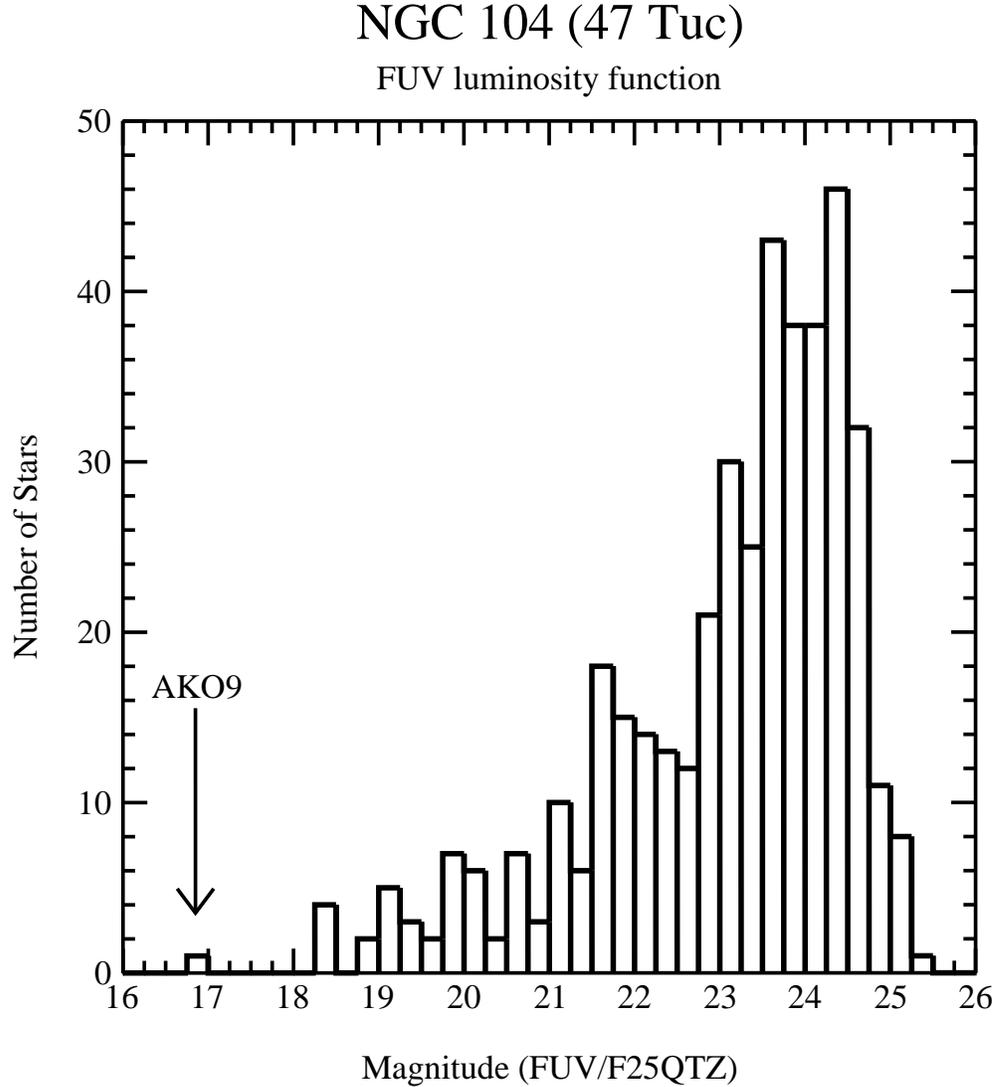}{400pt}{270.01}{75}{75}{-298}{450}
\caption{A preliminary FUV luminosity function (LF) derived from our
imaging data. This LF contains approximately 425 sources and is
complete to roughly $m_{\sc fuv} \simeq 24$. The brightest
FUV source -- AKO~9 -- is marked. FUV magnitudes are given in the
STMAG-system, i.e. $m = -2.5 \log{F_{\lambda}} - 21.1$, where
$F_{\lambda}$ is the flux of a source with a flat spectrum in the
relevant pass band.} 
\end{figure}

Figure~5 shows a first FUV-optical color-magnitude diagram (CMD) we
have constructed from our data. Given the large number of sources in
the optical images, it is not trivial to match FUV sources with
optical counterparts. As a result, the preliminary CMD in Figure~5
only includes the 50 or so objects for which this was relatively
straightforward. Note that this explains the lack of sources on or
close to the WD sequence in the CMD: optical counterparts for such
optically faint objects are hard to find. We stress that this is a
temporary state of affairs. Deep and well-resolved optical images of
the cluster core are already available (Gilliland et al. 2000), so we
fully expect to find optical counterparts for most or all of our FUV
sources.

\begin{figure}
\plotfiddle{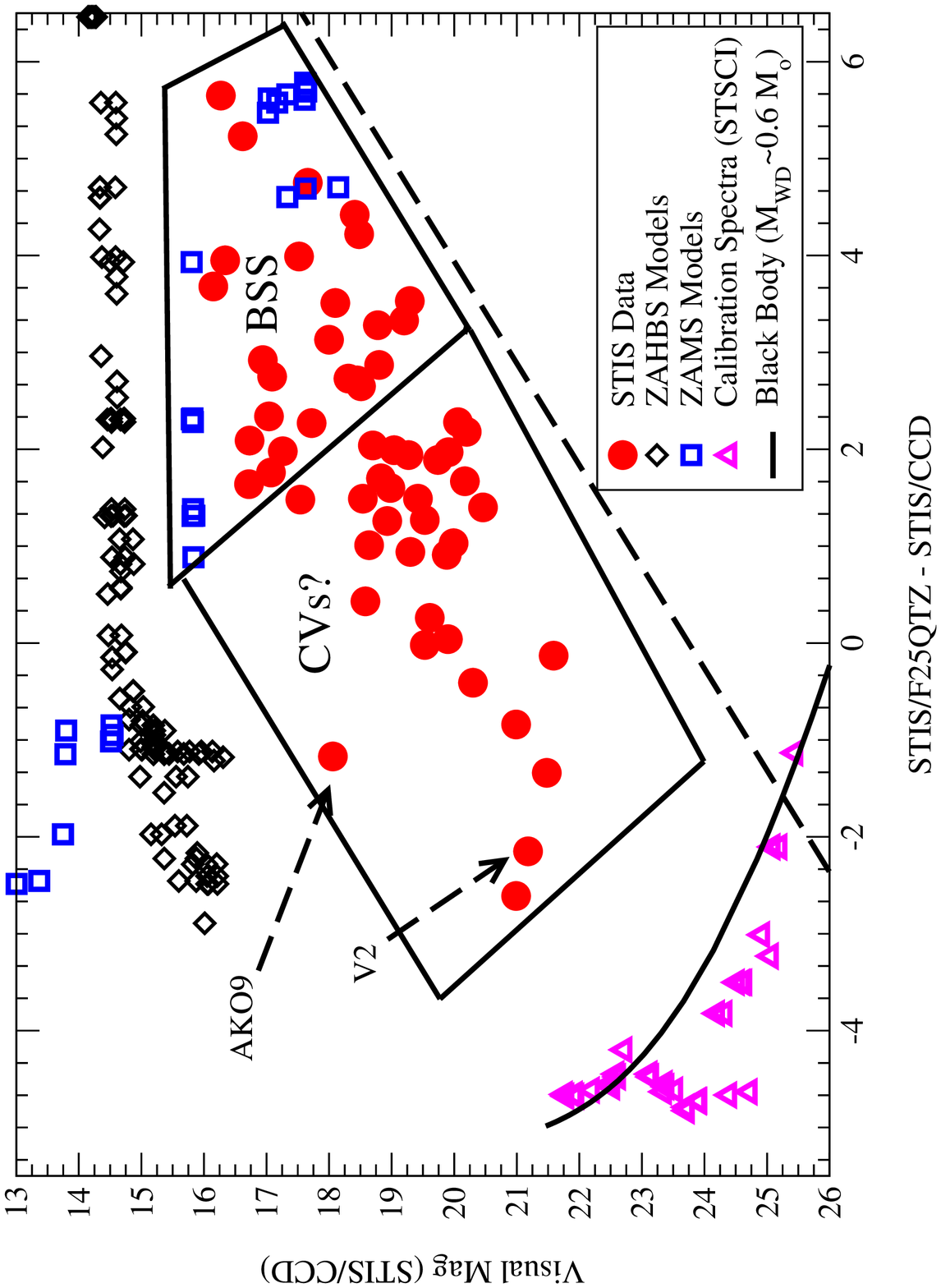}{405pt}{270.01}{70}{70}{-275}{430}
\caption{ A preliminary FUV-optical color magnitude diagram
constructed from our STIS observations. The new data points are shown
as filled circles. At the moment, only FUV sources
with relatively obvious optical counterparts are included in this CMD.
The dashed diagonal line marks the detection limit of the FUV data. 
The open triangles show the expected position of the WD
sequence, as obtained from synthetic photometry (with {\sc synphot}) of
STIS calibration spectra of field WDs. A simple blackbody WD sequence
is also shown. 
The open squares mark the position of the MS extension. These points 
have been calculated from the library of zero-age MS models in {\sc
synphot}. The actual MS in 47~Tuc ends roughly at the intersection 
of the ZAMS models and the FUV detection limit. The open diamonds mark
the expected position of the horizontal branch (HB), as
calculated from the library of zero-age HB models in
{\sc synphot}. The 20 or so STIS sources closest to the MS extension in
this plot are likely to be blue stragglers. The remaining 30 or so 
sources are located between the WD and main sequences and are
therefore 
candidate CVs.} 
\end{figure}

Luckily, the limited number of objects that have made it into Figure~5
are extremely interesting. The 20 or so stars closest to the MS
extension are most probably blue stragglers (BSs), i.e. the
rejuvenated products of stellar collisions or mergers (Livio 1993). The
existence of a BS population in the core of 47~Tuc is well established
(Paresce et al. 1991), but our program will be the first to provide
FUV spectra for a representative sample of BSs.

The remaining 30 or so objects in Figure~5 are even more interesting.
They are not part of either the WD sequence or the extended
MS, and instead occupy a region squarely between the two. This is
precisely the area of the CMD that should be populated by CVs. After
all, CVs contain both a UV bright, accreting WD and an optically
bright MS companion. If this identification is correct, then we have
finally discovered the large population of faint GC CVs whose
existence is predicted by tidal capture theory. 

It is reassuring that the only confirmed CV in 47~Tuc, the DN V2
(Shara et al. 1996), is 
among the CV candidates in our FUV-optical CMD. However, it is also
interesting to note that V2 and AKO~9 (see Section~3.2) are the only
two FUV sources detected so far that exhibit obvious emission lines in
the FUV spectral image (Figure~3). Is it possible that this is just a
selection effect, i.e. are most of our CV candidates simply to faint
to reveal their emission lines in Figure~3? Or does it mean the
remaining sources are not, in fact, CVs at all? Or perhaps GC CVs are
different from field CVs in presenting much weaker lines, perhaps due
to the lower metallicity in the cluster environment? 

Clearly, much work is still needed to verify our potential discovery
of a large CV population in 47~Tuc. However, all of it can be done
with available data. First, it is important to identify {\em all} CV
candidates in our FUV images, not just the optically brightest
ones. This simply means that we need to push both the FUV and optical
photometry deeper and carefully match the two. Second, we need to
search for time variability among the CV candidates. As noted above,
once all of our HST visits have been analyzed, we will be able to look
for variability on a wide range of time scales, from tens of minutes
(flickering) to hours (orbital variations) to months (DN
eruptions). Third, we obviously still need to extract the spectra of
the FUV sources from our slitless spectral images. The immense value
of these spectra is best illustrated by an example, so we now turn our
attention to the brightest FUV source in our data, AKO~9.

\subsection{New Light on AKO~9}

The nature of AKO~9 has been a long-standing puzzle: it is known to be
a eclipsing binary with an orbital period of about one day (Edmonds et
al. 1996); it has been suggested (Auri\`{e}re, Koch-Miramond \& Ortolani
1989; Geffert, Auri\`{e}re \& Koch-Miramond 1997) and refuted (Verbunt \&
Hasinger 1998) as the counterpart of a known x-ray source in the
cluster; it is clearly UV-bright (Auri\`{e}re, Koch-Miramond \&
Ortolani 1989; this paper); and 
it exhibits optical high and low states separated by approximately 2-3
mag (Minniti et al. 1997). All of these clues have yet to be assembled
into a convincing interpretation.

We begin our new look at AKO~9 by considering what can be learned from
the available optical data. Figure~6 shows the phase-folded V- and
I-band light curves of the system, as extracted from Gilliland et al.'s
(2000) observations of 47~Tuc. Two important features are evident: (i)
a weak primary eclipse is clearly present in the V-band, but is barely
noticeable in the I-band; (ii) both optical light curves are
double-humped.

\begin{figure}
\plotfiddle{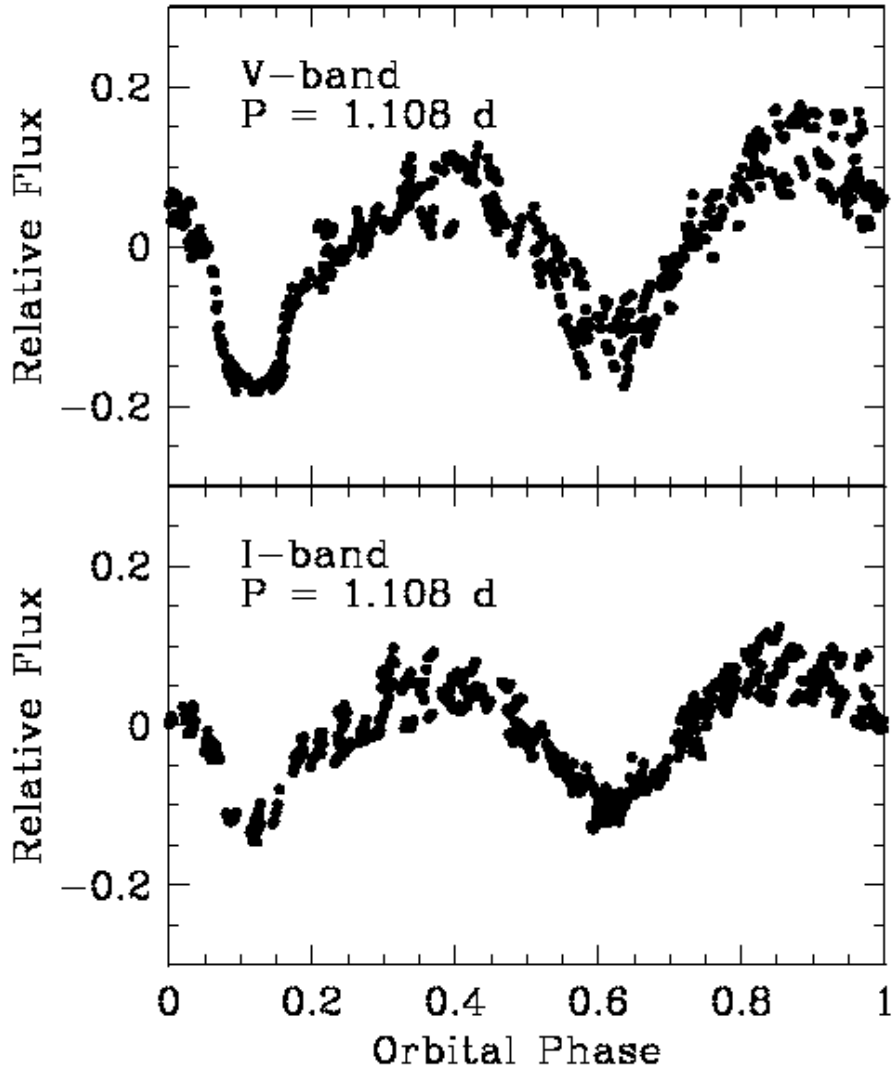}{500pt}{0}{70}{70}{-225}{-60}
\caption{The V- and I-band light curve of AKO~9 as derived from
Gilliland et al.'s (2000) HST/WFPC2 observations of 47~Tuc. The data
is shown folded onto the orbital period of 1.108~days. This period was 
derived from the data itself and is consistent with the 1.16~day 
period derived by Edmonds et al. (1996).}  
\end{figure}

The decreasing depth of primary eclipse with increasing wavelength
indicates that the spectrum
of the primary is much bluer than that of the secondary. In fact, a
look at AKO~9's U-band light curve (Figure~6 in Edmonds et al. 1996)
shows that at this wavelength the primary eclipse is at least 1 mag
deep. We can therefore conclude that the secondary dominates at both V
and I, whereas the primary dominates at U. The double-humped light
curve shapes are most easily interpreted as ellipsoidal variations
caused by the distorted shape of a Roche-lobe-filling secondary.  They
therefore strongly support the idea that AKO~9 is an {\em interacting}
binary system.

The amplitude of ellipsoidal variations depends on mass ratio
($q=M_2/M_1$) and inclination ($i$) and can be used to constrain these
parameters. In AKO~9, a second constraint comes from the presence of
primary eclipses. In a system containing a Roche-lobe-filling
secondary, the size of the secondary relative to the binary
separation, $R_2/a$, is only a function of $q$. We can therefore set a
lower limit on inclination as a function of mass ratio, $i >
\cos^{-1}{R_2/a}$. Figure~7 shows the allowed parameter space for
AKO~9 in the $q-i$ plane. Overall, the limits are $q < 1.8$ and $i >
69^o$.

\begin{figure}
\plotfiddle{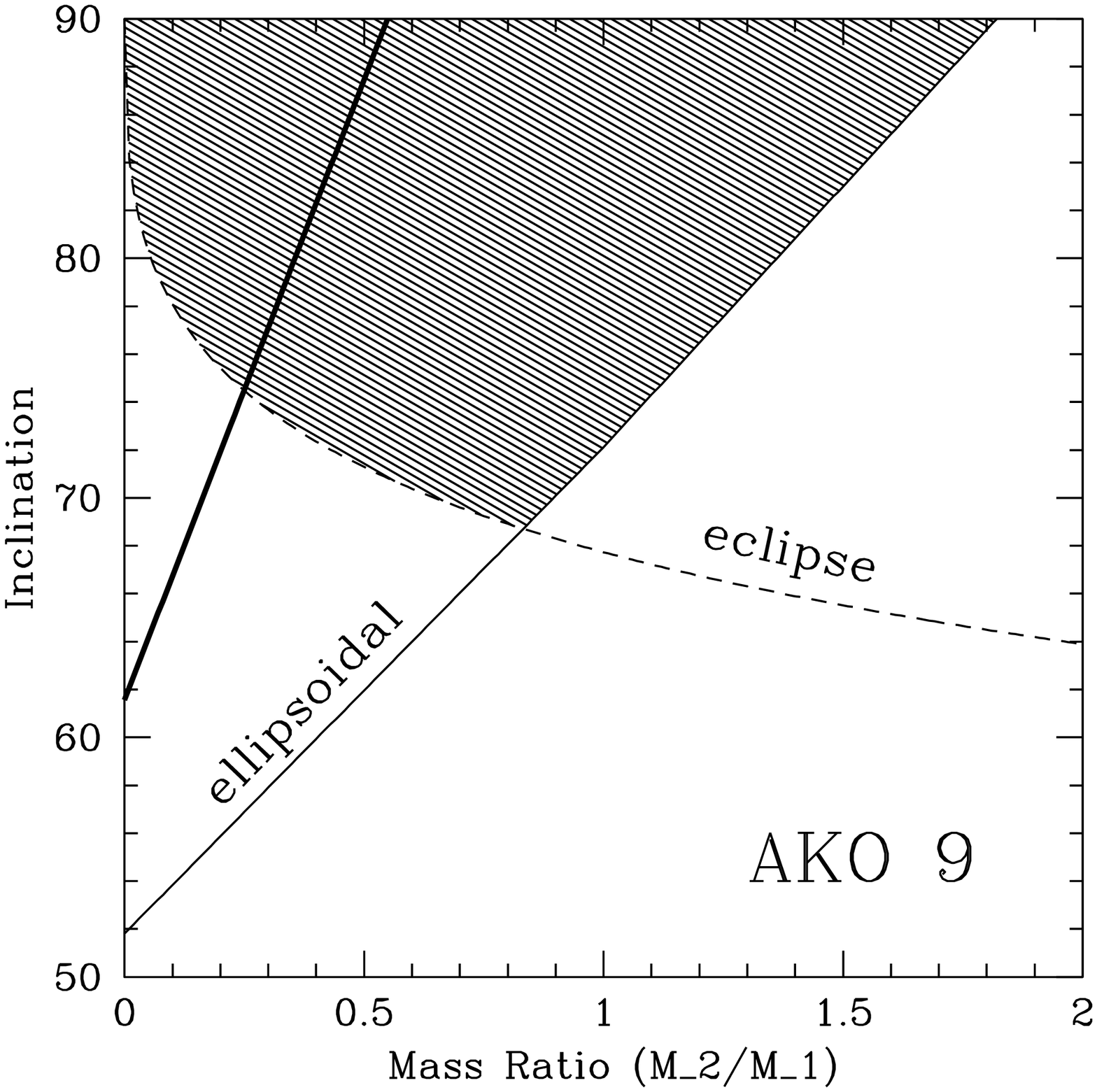}{450pt}{0}{70}{70}{-225}{-70}
\caption{Constraints on AKO~9's mass ratio and inclination. The two
solid lines are based on the amplitude of the ellipsoidal variations: 
the thick line is the most likely $q$ vs $i$ relation, the thin line
marks the lower limit on $i$. The dashed line marks a second lower
limit on $i$, based on the fact that AKO~9 is an eclipsing system. The
shaded area is the allowed region in the $q$-$i$ plane.
The amplitude of ellipsoidal variations as a function of $q$ and $i$
was calculated by interpolation from the tables of Bochkarev,
Karitskaya \& Shakura (1979). For the calculation of gravity-
and limb-darkening coefficients we assumed that the secondary star is
on 47~Tuc's sub-giant branch. We therefore used the 13~Gyr isochrone of Hesser
et al. (1987) to estimate the secondary's effective temperature and
surface gravity and then took the appropriate coefficients from
Bochkarev et al. (1979) and Diaz-Cordov\'{e}s, Claret \& Gim\'{e}nez
(1995).}  
\end{figure}

The ellipsoidal variations also imply that the secondary star cannot
be on the MS. Combining Kepler's law, Eggleton's (1983) approximation
for the Roche-lobe radius and AKO~9's orbital period, we find that for
any reasonable MS mass-radius relation, the mass and radius of a
lobe-filling MS star would have to be over 3 times solar, regardless
of the mass ratio of the system. Such a star should have evolved off
the cluster MS long ago (the MS turn-off occurs at about $M_{to}
\simeq 0.9 M_{\odot}$ in 47~Tuc; Hesser et al. 1987). The donor star
must therefore be an evolved object. This is consistent with AKO~9's
V-band magnitude, which corresponds to the sub-giant branch of the
cluster (Hesser et al. 1987).

All of this suggests that mass transfer in AKO~9 has started
relatively recently, as a direct consequence of the nuclear evolution
of the secondary. More specifically, mass transfer was probably
initiated when the radius of the donor star caught up with the Roche
lobe during the donor's evolution from the MS to the RGB via the
sub-giant 
branch. If this scenario is correct, the mass of AKO~9's secondary
must still be roughly that of an isolated star on 47~Tuc's sub-giant
branch, i.e. $M_2 \simeq 0.9 M_{\odot}$ (Hesser et al. 1987). We can
combine this with the upper limit on the mass ratio to obtain a lower
limit of $M_1 > 0.5 M_{\odot}$ for the mass of the primary.

We now turn our attention to the appearance of AKO~9 in our FUV
data. Figure~8 shows two cropped FUV images of the vicinity of AKO~9.
Both images were taken during the 1999 pilot visit, the first near the
beginning, the second near the end (about 7 hrs later). It is
immediately obvious that AKO~9 faded dramatically over the course of
this visit. We initially thought that we might have caught the system
on the decline from one of its bright states, but the correct
explanation turns out to be more mundane. Figure~9 shows AKO~9's light
curve over the course of the visit, as derived from our FUV
spectroscopy (this provides much better time resolution than the FUV
photometry). The light curve shape clearly suggests that our
observations coincided with the ingress into primary eclipse. The
timing of the observed fading is consistent with this interpretation:
it agrees very well with the time of eclipse predicted by a rough
orbital ephemeris we were able to derive from Gilliland et
al.'s (2000) optical data.

\begin{figure}
\plottwo{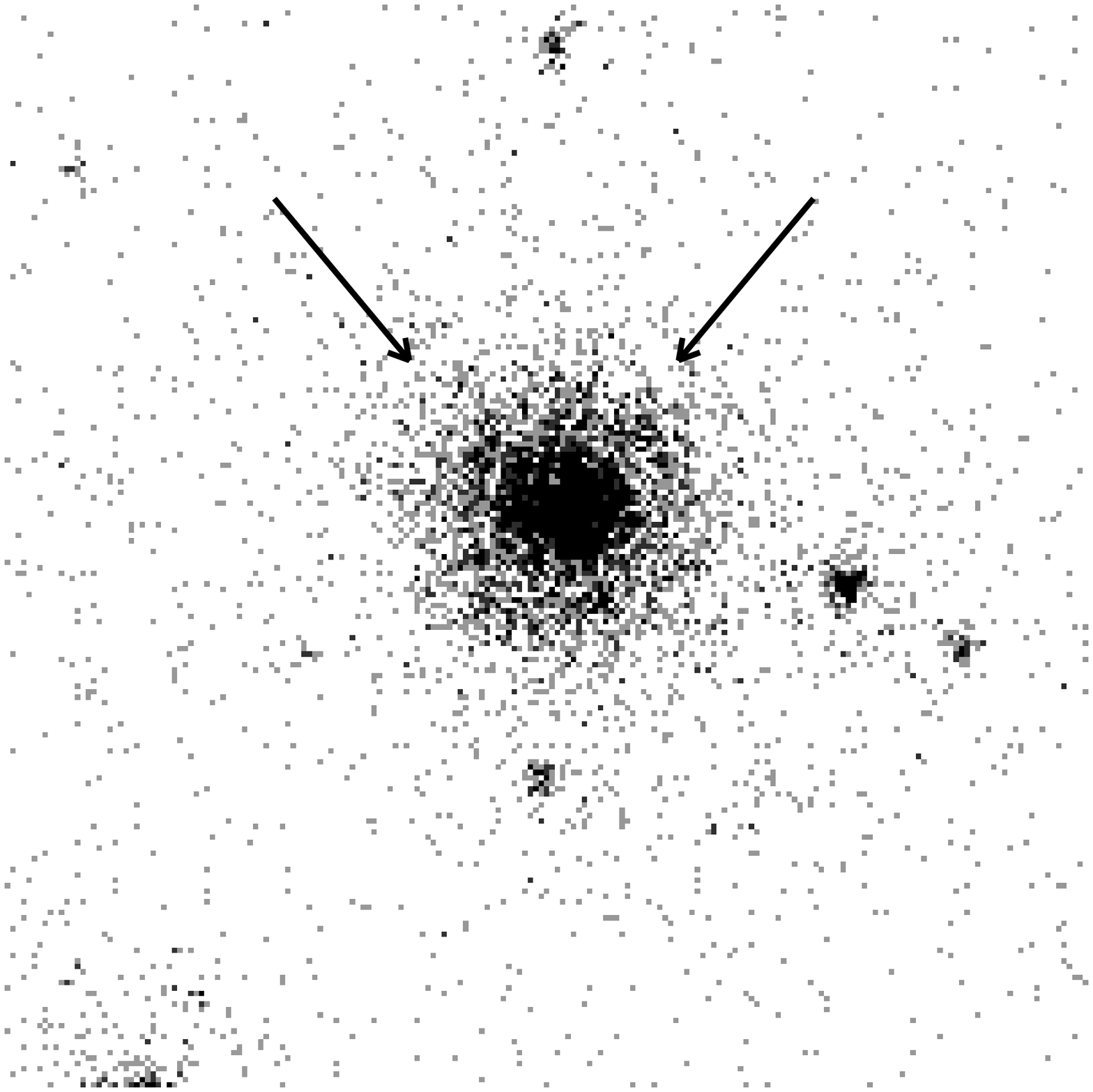}{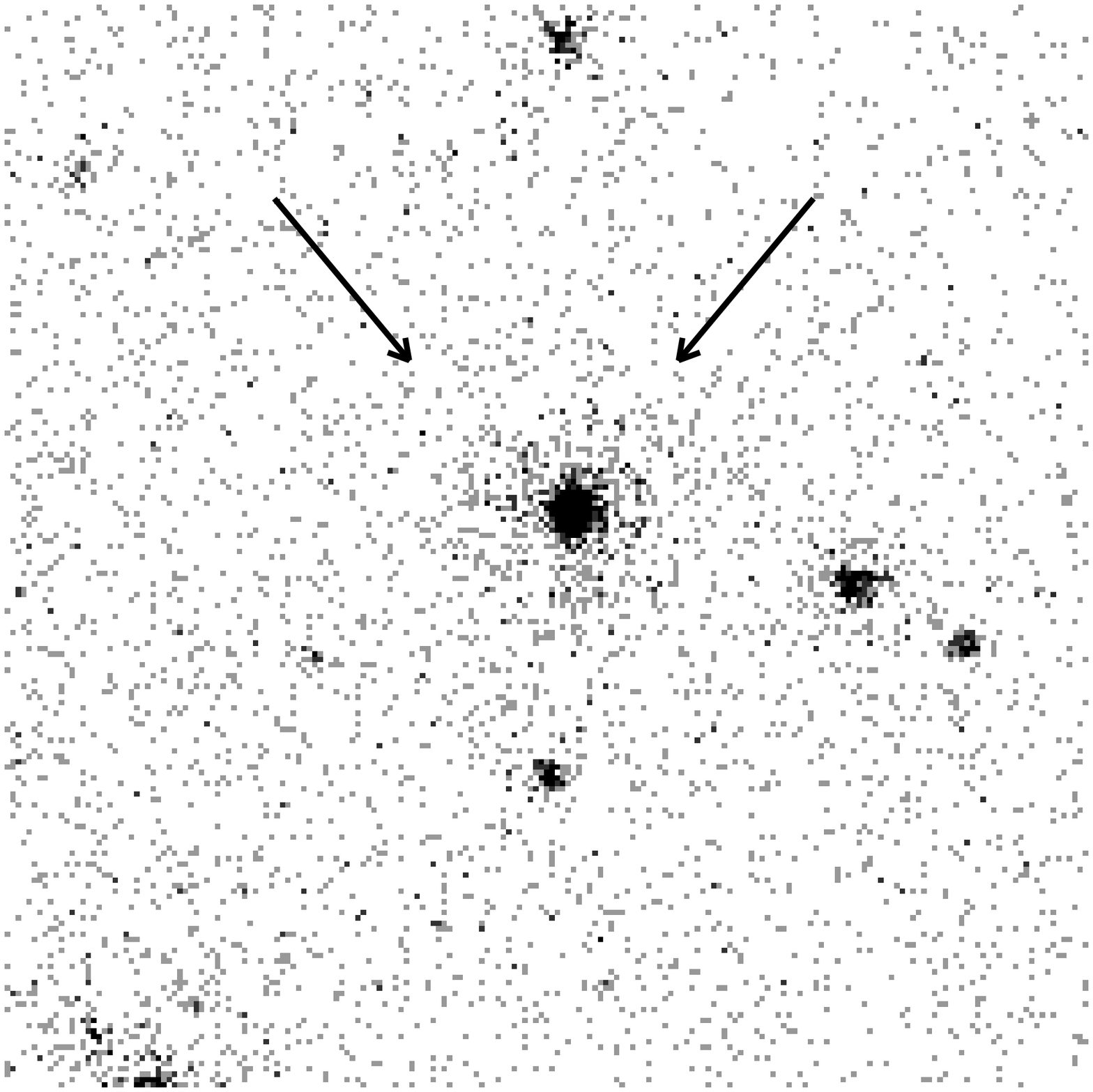}
\caption{Two cropped FUV images of the vicinity of AKO~9. The 
first image was taken early on during the 1999 HST visit, the second 
approximately 7 hours later near the end of the same visit. AKO~9
faded dramatically between these two exposures.}
\end{figure}

\begin{figure}
\plotfiddle{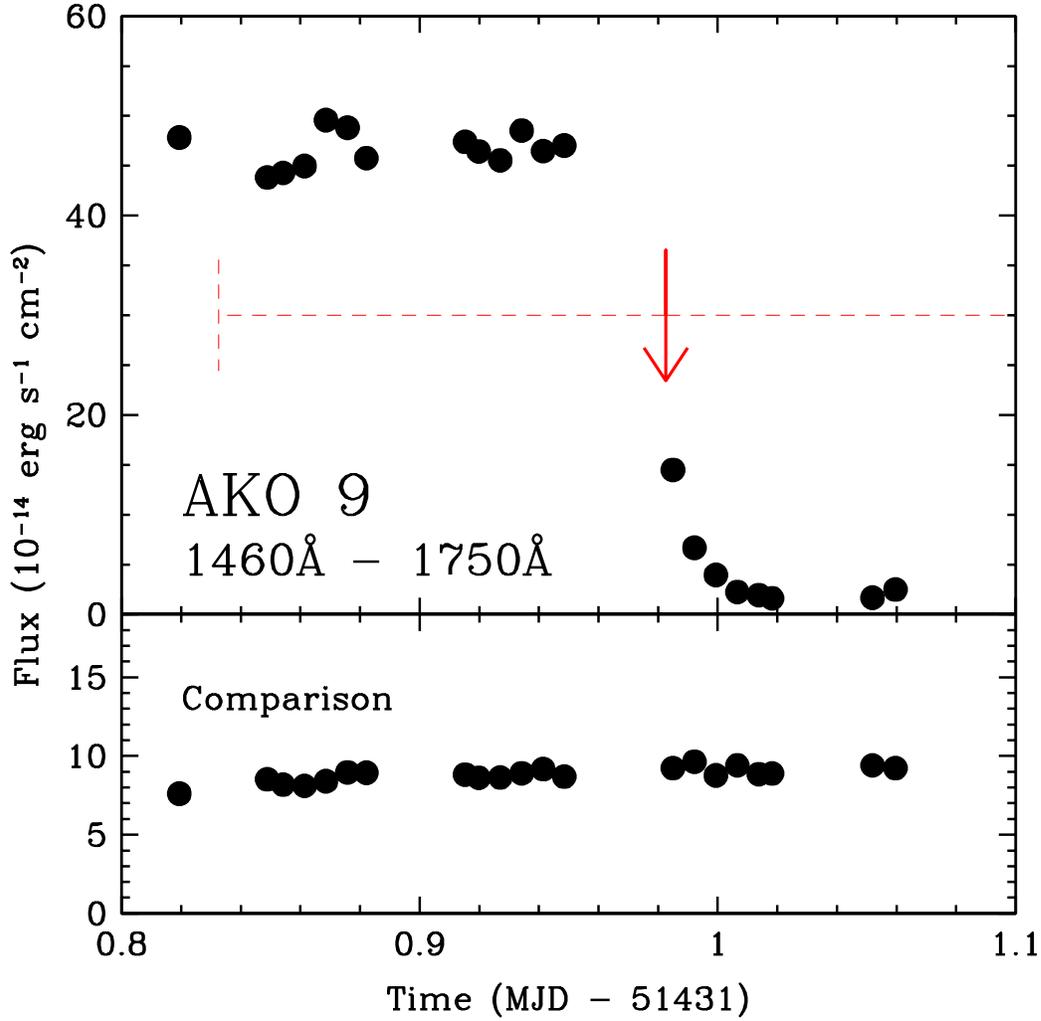}{380pt}{0}{70}{70}{-225}{-110}
\caption{The FUV light curve of AKO~9 during our 1999 HST visit, as
derived from the slitless spectroscopy. The arrow marks the expected
position of the eclipse; the horizontal line is the error on this
prediction. AKO~9's fading is clearly consistent with being caused by 
the ingress into the eclipse. The bottom panel shows the light curve
of another FUV bright source in the same data set.}
\end{figure}

Figure~10 shows the FUV spectrum of AKO~9 and illustrates its
evolution through the eclipse. The spectra in Figure~10 are based on
somewhat preliminary, quick-and-dirty extractions from our slitless
spectral images, but their overall spectral shapes and strong features
should be reliable. The pre-eclipse spectrum is very blue, with strong
C~{\sc iv} and He~{\sc ii} emission lines. A power law fit to the
dereddened continuum ($F_{\lambda} \propto \lambda^{-\alpha}$) yields
a spectral index of $\alpha=2.85 \pm 0.15$. Extrapolating this fit to
the U-band predicts $U \simeq 18.7 \pm 0.1$. This is approximately
$\Delta m_U \simeq 0.7 \pm 0.1$~mag fainter than the U-band magnitude
of AKO~9 in its normal, low state. The power law extrapolation should
not be taken too seriously, but it is interesting to note that $\Delta
m_U$ is reasonably consistent with the presence of 1-mag U-band
eclipses. More specifically, if the U-band light at mid-eclipse is
completely due to the secondary, the primary should be approximately
$\Delta m_U \simeq 0.6$~mag fainter than the total out-of-eclipse
U-band magnitude.

\begin{figure}
\plotone{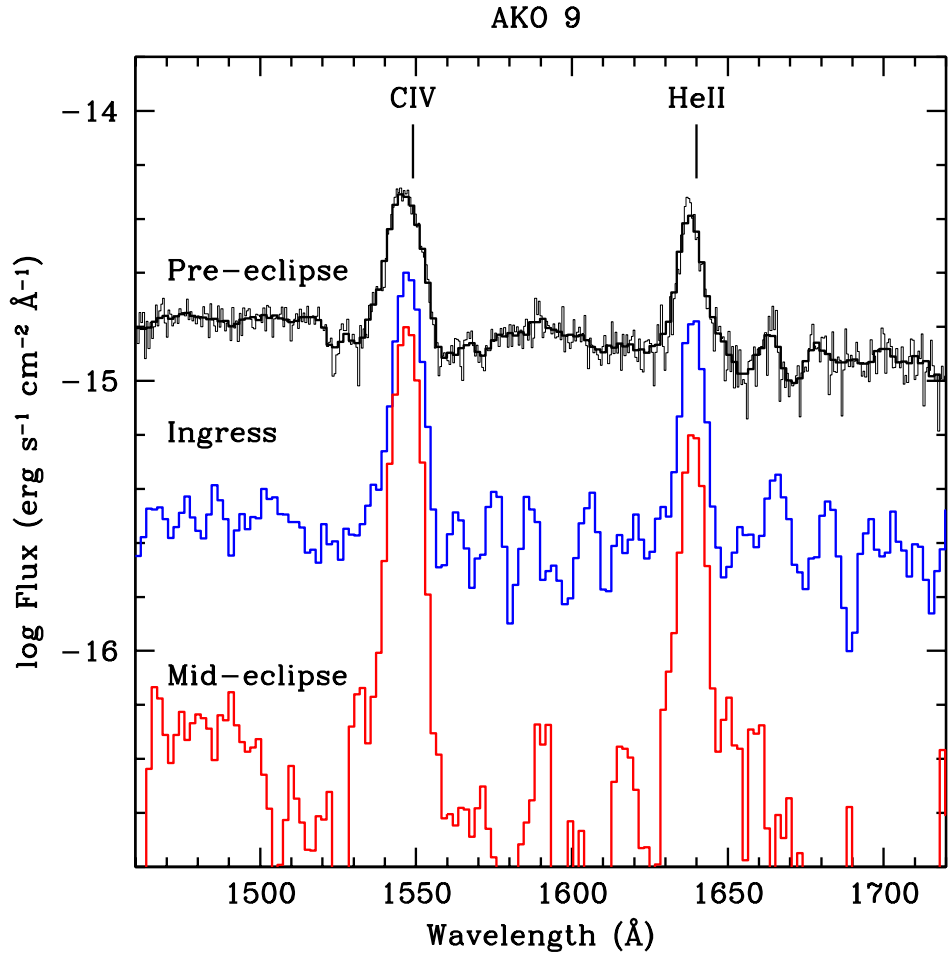}
\caption{The FUV spectrum of AKO~9 and its evolution through the 
primary eclipse. Note that the emission lines are much
more weakly eclipsed than the continuum. All spectra have been 
smoothed to increase their S/N, but the thin line superimposed on the
pre-eclipse spectrum shows the unsmoothed data for comparison.}
\end{figure}

The pre-eclipse FUV spectrum of AKO~9 looks remarkably like that of
certain CVs. This similarity extends to the fact that the strong
spectral lines are much more weakly eclipsed than the continuum. This
effect is common among field CVs and is generally ascribed to line
formation taking place in a vertically extended region, such as an
accretion disk wind (e.g. Knigge \& Drew 1997). AKO~9's 1.1~d orbital
period is, of course, rather long for a CV. It is therefore important
to compare its properties specifically to those of long-period
field CVs. Such a comparison is shown in Figure~11, where AKO~9's FUV
spectrum is compared to that of the well-known, long-period CV
GK~Per. The similarity between the two spectra is obvious.

\begin{figure}
\plotone{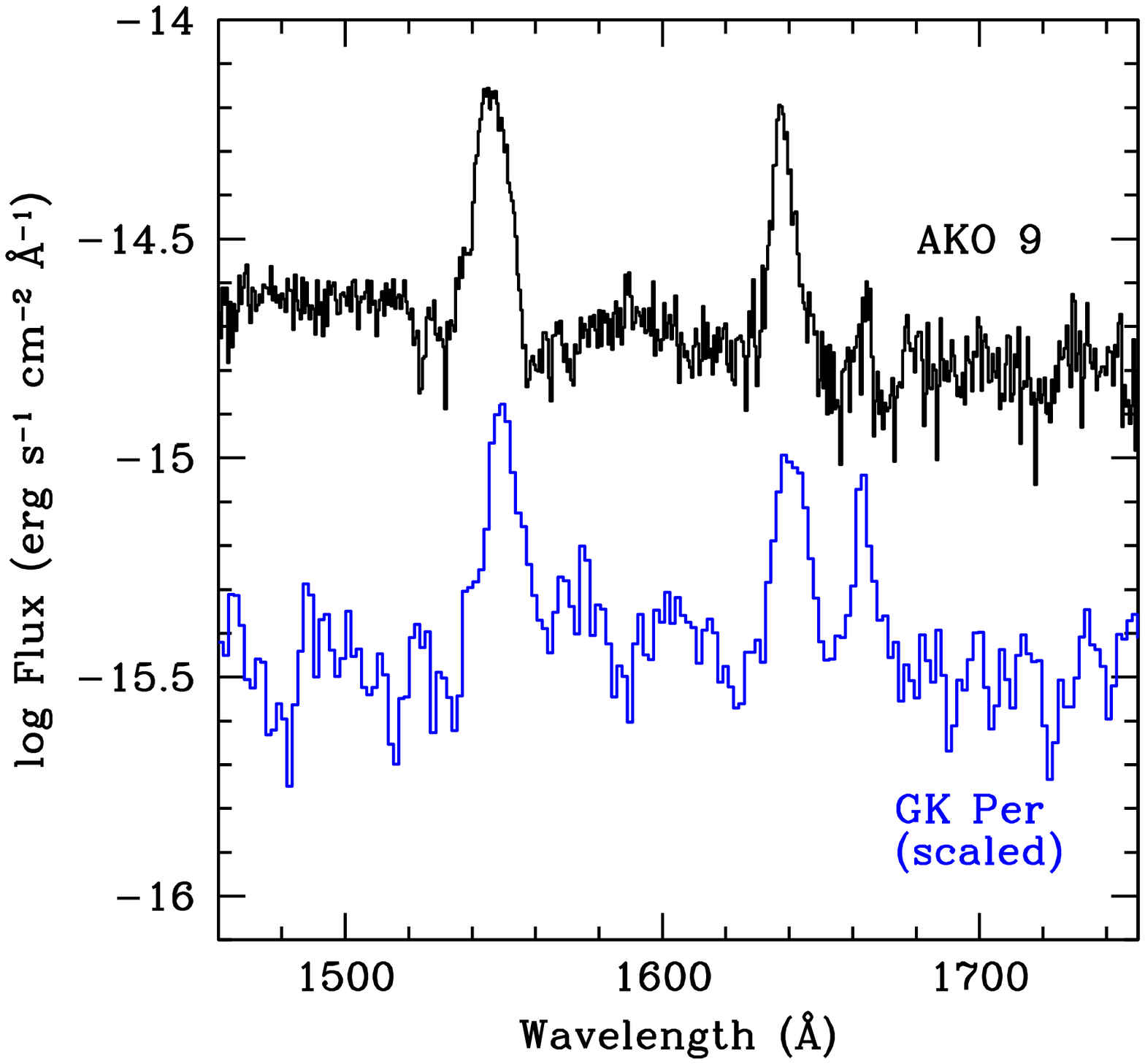}
\caption{The FUV spectrum of AKO~9 (pre-eclipse) compared to that of
the long-period field CV GK~Per. The spectrum of GK~Per has been
extracted from the IUE archive and scaled to the distance and
reddening of 47~Tuc. The overall appearance of the spectra is
similar. The three strongest spectral lines are C~{\sc iv}~1550~\AA,
He~{\sc ii}~1640\AA~and (probably) O~{\sc iii]}~1660~\AA.}
\end{figure}

GK~Per is thought to be an intermediate polar, in which a magnetic WD
accretes from an evolved sub-giant companion via a magnetically
truncated accretion disk. Its orbital period is approximately 2~days,
and it is known to undergo DN eruptions. If the similarities between
GK~Per and AKO~9 extend to CV sub-type, AKO~9's occasional high states
probably represent DN outbursts. Minniti et al. (1997) inferred a
roughly 1~hr rise time for the low-state-to-high-state transition from
a rapid brightening of the system in their U-band HST
observations. This would seem to argue against a DN interpretation,
since the rise times of DN eruptions are generally much
longer. However, we suspect that the brightening observed by Minniti
et al. was not a true low-state-to-high-state transition, but instead
an egress from eclipse at a time when AKO~9 was already in a high
state. Minniti et al. apparently discarded this possibility because
the brightness increase they observed was in excess of 2~mags --
considerably larger than the 1~mag eclipse depth seen in the low
state. However, if high states are caused by increases in accretion
luminosity, the fractional contribution of the accreting primary to
the total light will be much larger in the high state than in the low
state. The primary eclipse should therefore be expected to be much
deeper in the high state.

Considering all of the available evidence, it seems very likely that
AKO~9 is a DN-type CV in which mass transfer is driven by the nuclear
evolution of a sub-giant donor star. As we have shown, all of AKO~9's
observational characteristics can find a natural interpretation within
this framework. We therefore consider AKO~9 to be the first new,
spectroscopically confirmed GC CV to come out of our survey. Given the
results presented in the previous section, many more may follow.

\section{Summary and Conclusions}

We have presented first results from a FUV spectroscopic and
photometric survey of the GC 47~Tuc. Standard tidal capture theory
predicts that well over 100 faint CVs should presently reside in
47~Tuc (di Stefano \& Rappaport 1994). Our main goal is to either 
confirm or rule out the existence of such a large CV population in 
the cluster core. 

We have so far analyzed only a fraction of the data, but
have already detected approximately 425 FUV sources. Most of these are
probably hot, young WDs, but we have also identified approximately
30 FUV sources whose position in the FUV-optical CMD makes them strong
CV candidates. If most of these objects are confirmed to be CVs by a
careful analysis of the full data set, tidal capture theory will have
been vindicated. 

Our observations have also allowed us to resolve the long-standing 
puzzle surrounding AKO~9, a UV-bright and highly variable 1.1~day
binary system in 47~Tuc. AKO~9 is the brightest FUV source in our
data, and presents a blue FUV spectrum with strong C~{\sc iv} and He~{\sc ii}
emission lines. Its spectrum is strikingly similar to that of the
long-period DN-type CV GK~Per. Based on this similarity and other
evidence, we suggest that AKO~9 is a long-period CV in which
mass-transfer is driven by the nuclear evolution of a sub-giant donor
star.

\end{document}